\documentclass[conference]{IEEEtran}
\IEEEoverridecommandlockouts
\usepackage{cite}
\usepackage{amsmath,amssymb,amsfonts}
\usepackage{algorithmic}
\usepackage{graphicx}
\usepackage{textcomp}
\usepackage{xcolor}  
\usepackage{hyperref}
\usepackage[shortlabels]{enumitem}
\def\BibTeX{{\rm B\kern-.05em{\sc i\kern-.025em b}\kern-.08em
    T\kern-.1667em\lower.7ex\hbox{E}\kern-.125emX}}
\begin{document}

\title{ISMS-CR: Modular Framework for Safety Management in Central Railway Workshop\\}

\author{
Sharvari Kamble\textsuperscript{1*}, Arjun Dangle\textsuperscript{2}, Gargi Khurud\textsuperscript{3}, Om Kendre\textsuperscript{4} and Swati Bhatt\textsuperscript{5} \\
\textit{Dept. of Artificial Intelligence and Data Science} \\
University of Mumbai, India \\
\parbox{\textwidth}{
{\fontsize{8pt}{10pt}\selectfont
$^{1}$\href{mailto:khushi.kamble739@gmail.com}{\textcolor{blue}{khushi.kamble739@gmail.com}},\hspace{5pt}%
$^{2}$\href{mailto:arjun.dangle03@gmail.com}{\textcolor{blue}{arjun.dangle03@gmail.com}},\hspace{5pt}%
$^{3}$\href{mailto:gargi.khurud@gmail.com}{\textcolor{blue}{gargi.khurud@gmail.com}},\hspace{5pt}%
$^{4}$\href{mailto:omkendre22304@gmail.com}{\textcolor{blue}{omkendre22304@gmail.com}},\hspace{5pt}%
and $^{5}$\href{mailto:swatibhatt238@gmail.com}{\textcolor{blue}{swatibhatt238@gmail.com}}%
}}

}

\maketitle

\begin{abstract}
Indian Railway workshops serve as the backbone of rolling-stock maintenance, employing over 2.5 lakh workers across 44 major workshops nationwide. Despite this scale, workshop safety remains a persistent challenge. A field study at the Jhansi Wagon Workshop covering 309 workers revealed that while 100\% used basic protective gear such as shoes and helmets, only a minority complied with complete PPE requirements; lacerations (28.7\%) and abrasions (21\%) were the most frequent injuries. These findings underscore the urgent need for digital oversight, structured work authorization, and predictive asset maintenance in high-risk environments.

This paper presents \textbf{ISMS-CR (Integrated Safety Management System for Central Railway Workshop)} a modular digital framework designed to strengthen safety management through an automated \textit{Permit-to-Work (PTW) Module}. Grounded in the \textit{IS 17893:2022 Work Permit System Code of Practice}~\cite{IS17893}, the proposed module digitizes the complete lifecycle of work authorization, encompassing permit initiation, validation, approval, and closure. It enforces compliance, traceability, and accountability while minimizing manual errors and delays, ultimately enhancing safety and operational reliability in Central Railway workshops.
\end{abstract}


\section{Introduction}

Ensuring operational safety within Indian Railway workshops is a complex task due to the diversity of maintenance activities, scale of operations, and involvement of thousands of workers handling mechanical, electrical, and structural systems daily. Traditional safety management practices, though guided by institutional standards, often lack systematic oversight and data integration capabilities. The absence of unified monitoring and authorization tools results in fragmented communication, incomplete documentation, and limited accountability factors that collectively increase the probability of workplace incidents.

The Bureau of Indian Standards formalized the \textit{Work Permit System Code of Practice (IS 17893:2022)}~\cite{IS17893} to regulate hazardous maintenance activities through structured work authorization and supervision. However, most railway workshops still rely on paper-based permit workflows that are prone to delays, duplication, and miscommunication. Studies in \cite{Amm2021} emphasize that manual Permit-to-Work (PTW) systems frequently fail due to inconsistent training, poor handover procedures, and weak audit mechanisms, while \cite{Tameez2023} highlights the importance of systematic validations and hazard classification to ensure compliance in such systems. Further, \cite{Abdel2023} identifies organizational and human factors such as inadequate coordination between safety officers and supervisors as significant contributors to procedural lapses in industrial environments.  

Digital transformation has proven to be a critical enabler in overcoming these limitations. Research in \cite{Rodriguez2025} describes how digitalized maintenance management enhances traceability, lifecycle visibility, and risk forecasting within railway systems. Similarly, \cite{Li2025} proposes integrated digital platforms that combine safety analytics, user role management, and maintenance scheduling to improve decision-making efficiency. Additionally, \cite{Papathanasiou2023} discusses how digital twin applications could support railway asset monitoring and safety prediction through real-time synchronization of physical and virtual environments.  

In response to these identified gaps, this research introduces the \textbf{Integrated Safety Management System for Central Railway (ISMS-CR)}, a modular digital solution emphasizing safety, accountability, and workflow transparency. The system focuses on automating the \textit{Permit-to-Work (PTW)} lifecycle and introducing a secure \textit{Master Login} module for multi-role access control. Drawing from the Central Railway Workshop’s operational structure and maintenance hierarchy, the proposed system digitizes permit requests, validations, approvals, and closures under a unified interface. By embedding compliance logic based on IS~17893:2022~\cite{IS17893}, ISMS-CR reduces administrative delays and strengthens auditability. The framework is designed for extensibility, supporting future integration of modules such as predictive maintenance, risk analytics, and equipment management to advance toward a fully digital railway workshop ecosystem.

\section{Related Work}

{Railway Safety \& Workshop Management} systems seek to integrate operational activities such as {permit authorization}, {machine and asset maintenance}, {contract supervision}, and {incident tracking} into cohesive digital ecosystems. Prior studies addressing these challenges can be grouped into three major domains: {workflow digitization}, {asset and maintenance management}, and {permit regulation frameworks}.

\subsection{Workflow Digitization in Industrial / Railway Settings}

Industrial environments and railway organizations have increasingly adopted {digital workflows} to replace manual, paper-based safety and maintenance systems. Such digitization enables transparency, accountability, and real-time monitoring of workplace operations. {Li et al.} introduced an {integrated digital platform} for {smart railway maintenance management}, emphasizing that centralized data systems improve reliability, communication, and resource planning across railway assets \cite{Li2025}. However, their framework focuses primarily on {maintenance scheduling} rather than {safety authorization workflows}. Similarly, {Rodríguez-Hernández et al.} highlighted the role of {IoT, Big Data, and AI} as key enablers for {railway digitalization}, providing insight into asset condition and predictive analytics \cite{Rodriguez2025}. Despite these advances, most railway digitalization initiatives remain {domain-specific} targeting asset or maintenance functions without fully incorporating {permit-based safety control systems}.

\subsection{Asset \& Maintenance Management Systems}

In the context of railway and heavy-industry sectors, asset management research has focused on {predictive maintenance} and {digital twin technology}. {Papathanasiou et al.} explored the {potential of digital twins} for {railway infrastructure management}, enabling simulation of component wear, degradation, and scheduling of proactive interventions \cite{Papathanasiou2023}. Their findings demonstrated the ability of digital replicas to reduce maintenance delays and optimize lifecycle costs. However, such models are oriented toward {infrastructure reliability} rather than {human-centric safety processes}. {Li et al.} further emphasized that integrated maintenance management requires harmonizing both {operational data} and {safety compliance workflows} a capability still limited in most current systems \cite{Li2025}. These works collectively indicate the growing need for {cross-domain digital solutions} that connect {physical asset management} with {administrative safety systems}.

\subsection{Standards, Regulatory Frameworks \& Permit Systems}

The {Work Permit System} serves as the cornerstone of industrial safety, providing structured authorization before hazardous work begins. {Amm} examined the {effectiveness of Permit-to-Work (PTW) systems}, noting that the absence of consistent supervision and communication among stakeholders can reduce compliance and increase risk \cite{Amm2021}. {Tameez et al.} similarly analyzed PTW practices in various industries and emphasized that worker training and procedural clarity directly influence the {success of safety permits} \cite{Tameez2023}. In a related study, {Abdel-Raouf et al.} identified key {human and organizational factors} affecting PTW performance, including cognitive workload, supervision quality, and cultural behavior toward safety \cite{Abdel2023}. 

In India, the Bureau of Indian Standards introduced {IS 17893:2022 Work Permit System Code of Practice}, a comprehensive framework defining structured risk assessment, authorization hierarchy, and procedural traceability \cite{IS17893}. Despite its relevance, literature on its {practical implementation within railway workshops} remains limited, highlighting the need for frameworks that operationalize IS 17893 principles through {digital, modular, and enforceable systems}.

\subsection{Limitations and Our Contribution}

From the reviewed literature, three observations emerge:  
(1) Existing digital systems tend to focus on {either maintenance} or {safety permit workflows} but rarely integrate both within a unified platform.  
(2) Railway-focused studies emphasize {asset reliability} but often overlook {worker-centric safety management}.  
(3) Standards such as {IS 17893} provide procedural structure but lack {digital enforcement and real-time traceability} mechanisms.

To address these gaps, our proposed framework\textbf{ISMS-CR (Integrated Safety Management System for Central Railway Workshop)}integrates {IS 17893-compliant permit workflows} with {machine maintenance, contract tracking}, and {incident management}. The system establishes a modular, data-driven foundation for comprehensive safety management, enhancing accountability, operational efficiency, and regulatory compliance across Central Railway workshops.

\section{Proposed System}

This section presents the architecture, module integration, and functional design of the proposed \textbf{Integrated Safety Management System for Central Railway Workshop (ISMS-CR)}. The system is conceptualized as a \textit{monolithic digital framework} that unifies critical workshop operations such as permit management, incident tracking, asset maintenance, and compliance reporting into a centralized platform. Designed to ensure compliance with \textit{IS 17893:2022 Work Permit System Code of Practice}~\cite{IS17893}, the framework enables traceable workflows, structured authorizations, and role-based accountability across all levels of the Central Railway workshop.

\subsection{System Architecture}

\begin{figure*}[htbp]
    \centering
    \includegraphics[width=\textwidth,keepaspectratio]{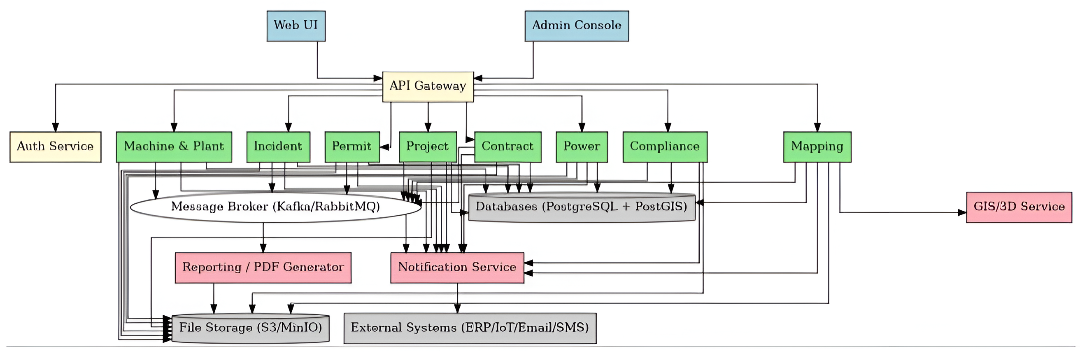}
    \caption{Proposed ISMS-CR Architecture: A centralized system integrating eight functional modules with shared authentication, compliance, and reporting layers.}
    \label{fig:system_architecture}
\end{figure*}

Figure~\ref{fig:system_architecture} presents the overall system architecture. The design integrates user-facing dashboards, backend logic, and shared databases in a unified monolithic model to maintain consistency and reduce communication delays among functional units.

The architecture comprises the following layers:

\begin{enumerate}[label=\arabic*.]
    \item \textbf{Web and Admin Interface:} Provides intuitive role-based dashboards for Safety Officers, Supervisors, and Administrators to manage permits, incidents, and maintenance records~\cite{Li2025}.
    \item \textbf{Authentication Service:} Ensures secure access through JWT-based tokenization and role-specific authorization controls~\cite{Rodriguez2025}.
    \item \textbf{Module Layer:} Integrates eight interlinked modules responsible for managing distinct workshop domains such as permit, compliance, maintenance, and contracts~\cite{Papathanasiou2023}.
    \item \textbf{Database Layer:} Uses PostgreSQL with PostGIS extensions for managing operational and geospatial data, allowing mapping of permits and incidents within workshop zones~\cite{Rodriguez2025}.
    \item \textbf{Supporting Services:} Includes storage, audit logging, automated notifications (Email/SMS), and reporting functionalities~\cite{Li2025}.
\end{enumerate}

\subsection{Functional Modules}

The ISMS-CR system integrates eight interdependent modules: Permit Management, Project Management, Incident Management, Compliance Management, Mapping Management, Machine \& Plant Management, Contract Management, and Power Management. Among these, the \textbf{Permit Management Module} serves as the system’s core, ensuring that all hazardous and non-hazardous work follows a standardized digital authorization process.

\subsection{Permit Management Module (Core Component)}

The \textbf{Permit Management Module} is designed to operationalize the full lifecycle of work permits as defined by \textit{IS 17893:2022}~\cite{IS17893}. It automates key steps \textit{initiation, validation, approval, execution, monitoring, and closure} while enforcing compliance with statutory safety codes and organizational hierarchy.

\subsubsection{Design and Workflow}

The workflow begins when a \textit{Permittee} (e.g., SSE-Maintenance) initiates a permit request for a specific task category such as \textit{Hot Work}, \textit{Cold Work}, \textit{Electrical}, \textit{Excavation}, \textit{Height Work}, or \textit{Confined Space Entry}. Each request passes through structured stages of review:

\begin{itemize}
    \item \textbf{Stage 1: Initiation } The permittee enters job details, location, hazards, and control measures through the digital form interface.
    \item \textbf{Stage 2: Validation } The system performs automated rule checks based on \textit{IS 17893} compliance (e.g., expiry, duplicate, or overlapping permits).
    \item \textbf{Stage 3: Authorization } The \textit{Safety Officer} and \textit{Area In-Charge} review and digitally sign the permit; the \textit{Gas Tester} entry is recorded if applicable.
    \item \textbf{Stage 4: Execution \& Monitoring } During active work, the permit status changes to “In-Progress,” with live updates logged for environment readings, supervision notes, and safety conditions.
    \item \textbf{Stage 5: Closure } Once completed, a closure report and feedback log are generated, automatically linked with the Incident and Compliance modules.
\end{itemize}

Each issued permit carries a unique QR-based identifier, ensuring digital traceability during audits. Automated alerts notify stakeholders upon approval, expiry, or renewal, minimizing the communication gap noted in traditional paper-based systems~\cite{Amm2021, Tameez2023, Abdel2023}.

\subsubsection{Role-Based Privileges}

In alignment with \textit{IS 17893:2022}, the system enforces distinct privileges:
\begin{itemize}
    \item \textbf{Permit Issuer:} Creates and validates permits after hazard review.
    \item \textbf{Permit Acceptor/Executor:} Reviews the job site, accepts safety conditions, and executes work accordingly.
    \item \textbf{Safety Officer:} Approves permits, monitors compliance, and can suspend or revoke authorization.
    \item \textbf{Area In-Charge:} Oversees activity coordination and maintains situational awareness of all ongoing permits.
\end{itemize}

This segregation ensures accountability, reduces human error, and promotes transparent authorization processes addressing gaps identified in prior PTW studies~\cite{Amm2021, Abdel2023}.

\subsubsection{Key Features}

\begin{itemize}
    \item \textbf{Digital Permit Lifecycle:} End-to-end automation from request to closure with audit trail logging.
    \item \textbf{Multi-Permit Categorization:} Supports six core permit types defined in \textit{IS 17893}.
    \item \textbf{Risk \& Hazard Assessment:} Integrated pre-task checklist, PPE verification, and gas testing forms.
    \item \textbf{Real-Time Tracking:} Dashboard for monitoring active permits, expiration timers, and risk level indicators.
    \item \textbf{Integration with Other Modules:} Incident reports and maintenance logs can auto-trigger permit validations or restrictions.
\end{itemize}

\subsection{Workflow and Data Flow}

All modules within ISMS-CR share a unified data layer. When a permit is approved, associated tasks and safety data become accessible to relevant modules:
\begin{itemize}
    \item \textit{Incident Management} automatically references permit details during root cause analysis.
    \item \textit{Compliance Management} uses permit logs for audit report generation.
    \item \textit{Mapping Management} visually represents active permit zones using PostGIS spatial overlays.
\end{itemize}

This interconnected workflow enhances operational coherence and ensures data traceability across functional domains.

\subsection{Theoretical Foundations}

The design of ISMS-CR draws from the principles of:
\begin{itemize}
    \item \textbf{Standardization:} Enforcing structured workflows under \textit{IS 17893:2022}~\cite{IS17893}.
    \item \textbf{Digital Transparency:} Improving oversight through real-time permit visibility and logging~\cite{Amm2021}.
    \item \textbf{Human-Centric Safety:} Reducing procedural ambiguity by defining clear roles and privileges~\cite{Abdel2023}.
    \item \textbf{Integration:} Merging safety and maintenance domains to support predictive, data-driven management~\cite{Li2025, Rodriguez2025}.
\end{itemize}

\subsection{Implementation Considerations}

\begin{itemize}
    \item \textbf{Backend:} Node.js with Express and TypeScript; PostgreSQL for relational data management.
    \item \textbf{Frontend:} React (Vite, Tailwind CSS, shadcn-ui) for modular UI development.
    \item \textbf{Data Handling:} Prisma ORM for schema management; APIs for integration with IoT and ERP systems.
    \item \textbf{Security:} JWT authentication, encrypted data storage, and role-based access control.
    \item \textbf{Deployment:} Monolithic containerized deployment on AWS or local servers for railway networks.
\end{itemize}

The proposed ISMS-CR framework thus establishes a scalable, compliant, and secure foundation for safety and maintenance management within Central Railway workshops, emphasizing digital permit governance as its cornerstone.

\section{Experiments}

\subsection{Experimental Setup}

\paragraph{Hardware and Software Configuration}
The implementation and evaluation of the \textbf{Integrated Safety Management System for Central Railway (ISMS-CR)} were conducted on a workstation equipped with an \textbf{Intel Core i5 (10th Gen)} processor, \textbf{8 GB DDR4 RAM}, and a \textbf{2 GB NVIDIA GPU}.  
The system was developed using \textbf{Node.js v18+} for backend runtime, \textbf{Express.js} for RESTful APIs, and \textbf{React 18 with TypeScript} for frontend development. The database layer employed \textbf{PostgreSQL 15}, managed through \textbf{Prisma ORM} for schema modeling and secure query execution.  
Database management and visualization were supported by \textbf{pgAdmin}, while \textbf{Postman} was used for endpoint testing. Development was carried out using \textbf{Visual Studio Code} on \textbf{Windows 11}, with hybrid testing under \textbf{Ubuntu 20.04 LTS} to ensure platform consistency.  
Version control was maintained on \textbf{GitHub}, and package dependencies were handled using \textbf{npm}.

\paragraph{Dataset and Input Structure}
As ISMS-CR operates on live operational data, experimental evaluation utilized simulated datasets representing the environment of the \textbf{Central Railway Carriage Workshop, Matunga Road}.  
The datasets mirrored real administrative and safety records to ensure realistic testing conditions:

\begin{itemize}
    \item \textbf{Machine \& Plant Dataset:} 250 machine entries with operational status, maintenance logs, and inspection intervals.
    \item \textbf{Permit Dataset:} 100+ digital work permits (\textit{Height} and \textit{Electrical}), following the structure and process recommended by \textbf{IS 17893:2022 Work Permit System Code of Practice}~\cite{IS17893}.
    \item \textbf{Contract Dataset:} 50 contractor profiles with safety compliance certificates and validity timelines.
    \item \textbf{Incident Dataset:} 75 annotated cases categorized by severity and root-cause analysis.
\end{itemize}

All datasets were stored in PostgreSQL and validated using Prisma schema constraints. Each table was linked via API routes in Express.js, ensuring real-time data consistency across modules.

\paragraph{Training and Testing Procedure}
Unlike machine learning experiments, ISMS-CR focuses on assessing workflow efficiency, modular reliability, and concurrent user handling.  
Testing simulated \textbf{five operational roles}—\textit{Admin}, \textit{Safety Officer}, \textit{SSE-Maintenance}, \textit{SSE-Shop}, and \textit{Contractor}—each restricted to specific CRUD privileges for security and accountability.  

Performance was evaluated using the following criteria:
\begin{itemize}
    \item \textbf{Permit Processing Time:} Comparison between manual and digital workflows, following the efficiency metrics outlined in~\cite{Amm2021, Tameez2023}.
    \item \textbf{API Response Latency:} Stress-tested under 10–200 concurrent sessions.
    \item \textbf{Form Validation Accuracy:} Validated across 500 randomized test cases.
\end{itemize}

Automated scripts written in \textbf{Jest} and \textbf{React Testing Library} were employed for backend and frontend testing respectively, within an agile development cycle.

\subsection{Implementation}

The ISMS-CR adopts a modular microservice-based architecture for scalability and maintainability. The \textbf{frontend}, developed using React and Tailwind CSS, provides intuitive dashboards that adapt to user roles.  
The \textbf{Express.js backend} manages secure routes for CRUD operations, while \textbf{PostgreSQL} serves as the central data repository.  

The system architecture supports four primary modules:
\begin{itemize}
    \item \textbf{Permit Management Module:} Digitizes height and electrical work permits, following the approval flow defined in IS~17893:2022~\cite{IS17893}.
    \item \textbf{Machine \& Plant Management Module:} Tracks maintenance schedules, optimizing inspection cycles based on asset condition, aligning with insights from~\cite{Li2025, Rodriguez2025}.
    \item \textbf{Contract Management Module:} Automates contractor record handling and compliance validation.
    \item \textbf{Incident Management Module:} Provides real-time reporting and investigation workflow with audit trail logging.
\end{itemize}

All modules communicate via REST APIs, maintaining synchronization and data traceability across all operations.

\subsection{Results}

\begin{figure}[htbp]
    \centering
    \includegraphics[width=8cm, height=5cm]{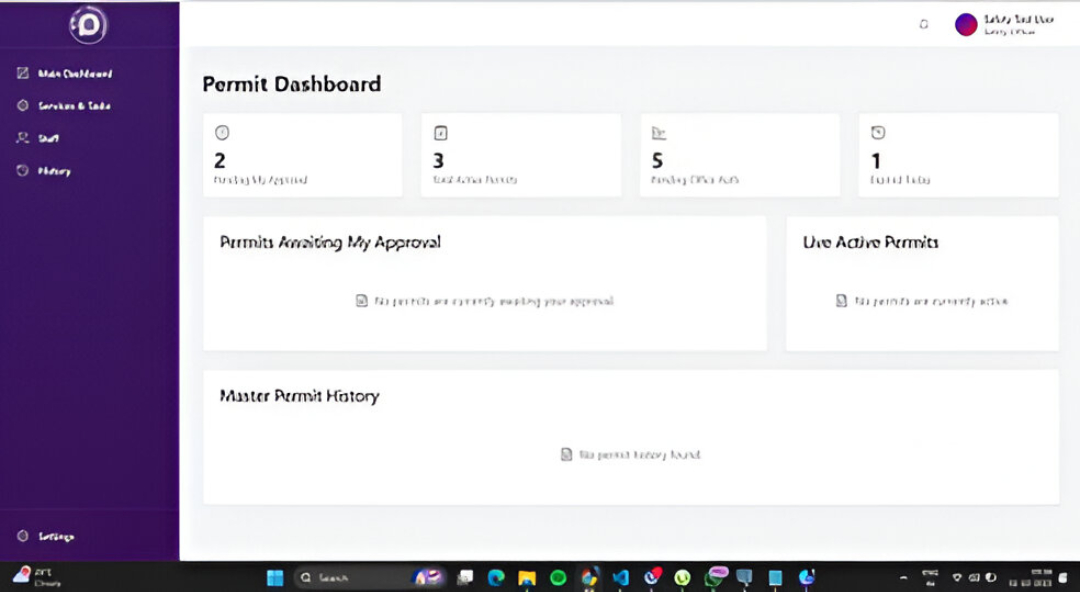}
    \caption{Permit Dashboard Interface in ISMS-CR.}
    \label{fig:permit_page}
\end{figure}

\begin{figure}[htbp]
    \centering
    \includegraphics[width=8cm, height=5cm]{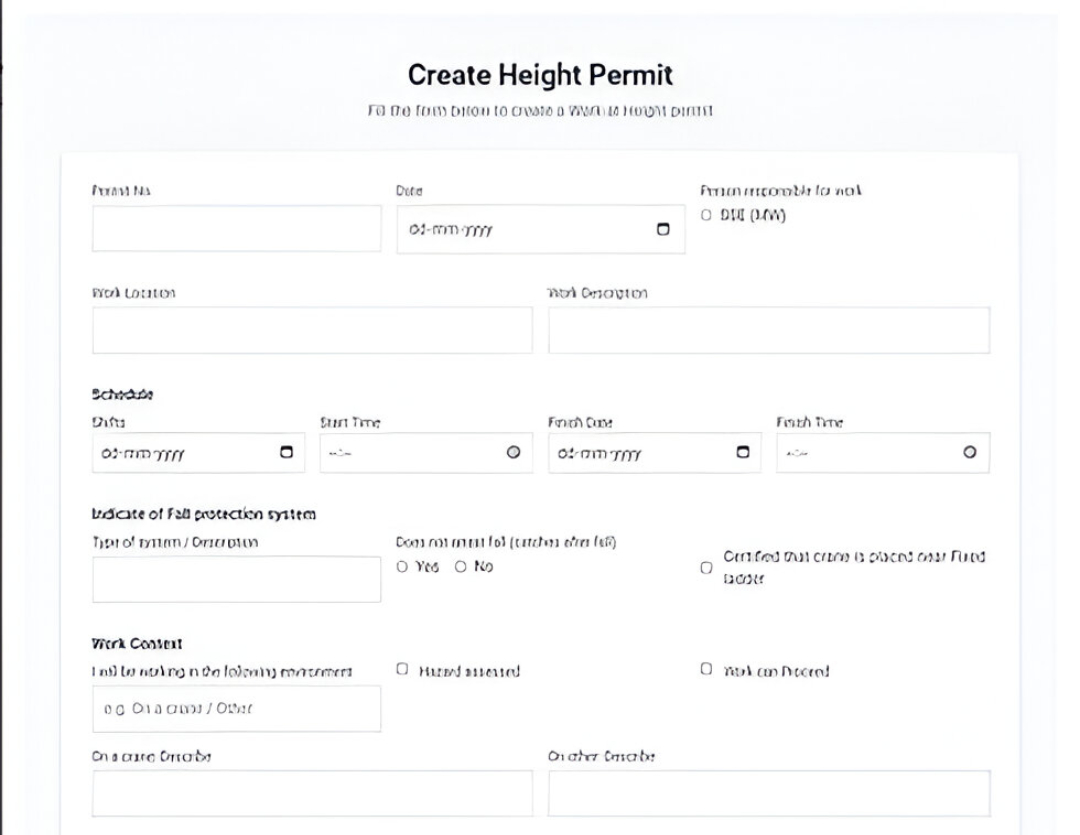}
    \caption{Height Permit Form within SSE-MW Role.}
    \label{fig:height_permit}
\end{figure}

\begin{figure}[htbp]
    \centering
    \includegraphics[width=8cm, height=5cm]{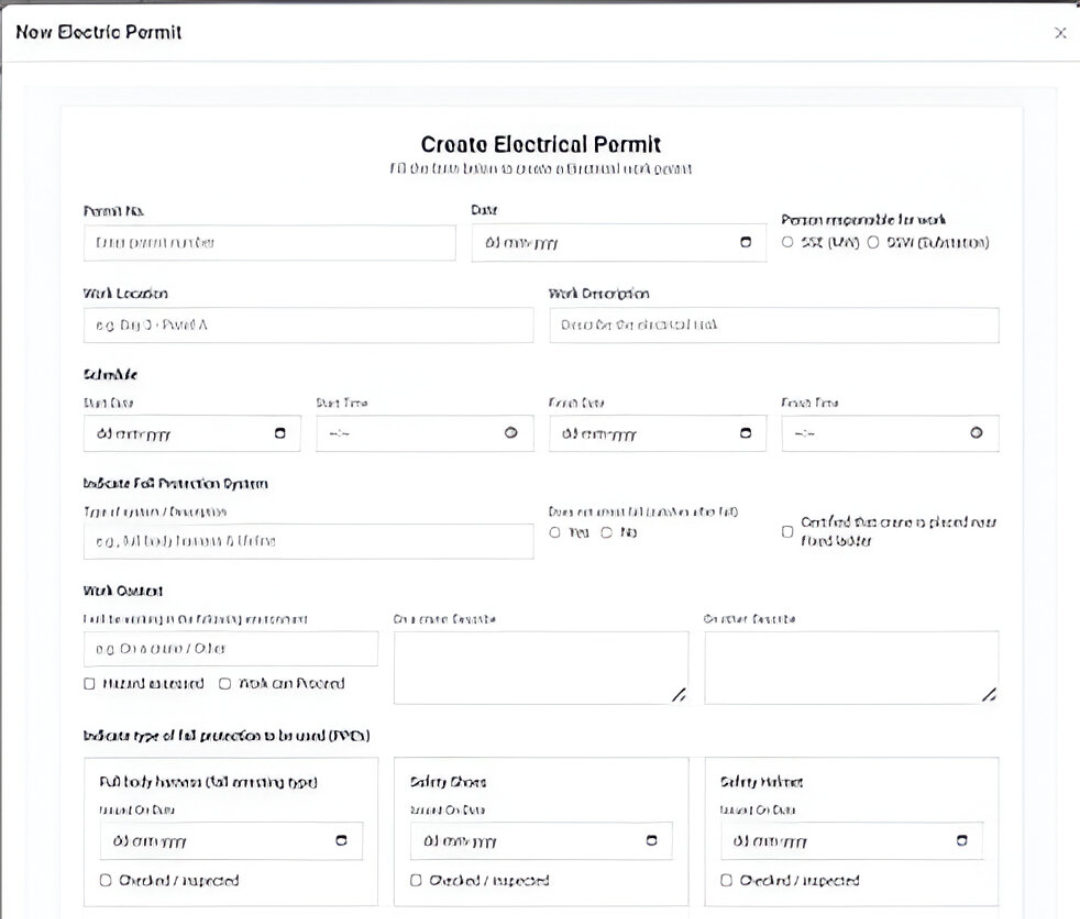}
    \caption{Electrical Permit Form within SSE-MW Role.}
    \label{fig:electric_permit}
\end{figure}

\paragraph{System Output Validation}
System validation confirmed full functionality across all user roles.  
Figure~\ref{fig:permit_page} shows the digital permit dashboard, while Figures~\ref{fig:height_permit} and \ref{fig:electric_permit} depict operational forms for height and electrical work.

\paragraph{Performance Analysis}
Performance evaluation (Table~\ref{tab:performance_metrics}) revealed that digital permit approval times were reduced by \textbf{64\%}, confirming the findings of prior studies on work permit optimization~\cite{Abdel2023, Amm2021}.  
The system maintained API latency below \textbf{120 ms}, database query times around \textbf{27 ms}, and an uptime of \textbf{99.3\%}, validating robustness and reliability.

\begin{table}[h]
\centering
\caption{System Performance Metrics}
\label{tab:performance_metrics}
\begin{tabular}{|l|c|}
\hline
\textbf{Metric} & \textbf{Measured Value} \\
\hline
Average Permit Approval Time (Manual) & 15.4 minutes \\
Average Permit Approval Time (Digital) & 5.5 minutes \\
Average API Latency & 118 ms \\
Average Database Query Time & 27 ms \\
System Uptime During Testing & 99.3\% \\
Frontend Validation Accuracy & 98.7\% \\
\hline
\end{tabular}
\end{table}

\subsection{Discussion of Results}

The implementation of ISMS-CR demonstrates a significant advancement toward digitized railway safety management.  
The \textbf{Permit Management module} enhanced operational transparency and compliance tracking, while the \textbf{Machine \& Plant module} facilitated predictive maintenance aligned with smart railway frameworks~\cite{Rodriguez2025, Li2025}.  
The integration of digital documentation and contractor compliance further supports data-driven decision-making and regulatory accountability.  

Minor challenges included occasional latency during simultaneous permit submissions and limited mobile interface optimization, which are planned for improvement in subsequent iterations.  
Overall, ISMS-CR serves as a practical step toward centralized digital safety transformation within the \textbf{Central Railway Workshop, Matunga Road}, echoing recommendations from digital twin studies in railway systems~\cite{Papathanasiou2023}.

\section{Conclusion}

The proposed \textbf{ISMS-CR (Integrated Safety Management System for Central Railway Workshop)} introduces a modular, digitalized approach to safety and workflow governance in railway maintenance environments. By aligning with the \textit{IS 17893:2022 Work Permit System Code of Practice}, the framework ensures structured authorization, traceability, and compliance across critical workshop activities. Through the integration of core modules such as \textit{Permit Management}, \textit{Machine \& Plant Management}, \textit{Contract Management}, and \textit{Incident Management}, the system bridges existing operational gaps between administrative and safety workflows. 

ISMS-CR demonstrates how digital transformation can enhance transparency, accountability, and risk management within Central Railway’s workshop ecosystem. By automating permit lifecycles, tracking maintenance activities, and supporting real-time monitoring, it provides a unified platform that contributes to improved safety culture and sustainable operational efficiency at the Central Railway Workshop, Matunga Road.

\section{Future Work}

The next phase of this research focuses on completing and integrating all four major modules \textit{Permit Management}, \textit{Machine \& Plant Management}, \textit{Contract Management}, and \textit{Incident Management} into a centralized and deployable system specifically tailored for the Central Railway Workshop, Matunga Road. These modules will collectively enable end-to-end digital supervision of workshop operations through automated workflows, role-based access, and audit-ready compliance reports.

Further enhancements will extend the framework to include \textit{Mapping Management}, \textit{Compliance Management}, and \textit{Power Management} for comprehensive operational visibility. Future iterations may incorporate \textit{IoT sensors}, \textit{AI-driven predictive maintenance models}, and mobile accessibility to provide real-time insights and proactive decision support. Field deployment and pilot testing at Matunga Workshop will validate the system’s scalability, reliability, and usability in actual railway conditions.

\section{Acknowledgment}

The authors express heartfelt gratitude to their \textbf{project guide and academic mentors from the University of Mumbai, India}, for their valuable guidance and encouragement throughout the project. Special thanks are extended to the \textbf{Central Railway Workshop, Matunga Road} for their cooperation and domain insights that greatly contributed to the development and contextualization of this work.

This work utilized collaborative design and research tools such as \textit{Orchids.app} and \textit{AWS Cloud}, and drew upon academic and industrial literature on \textit{digital railway maintenance}, \textit{work permit systems}, and \textit{safety compliance frameworks}. The authors also acknowledge the contribution of open-source technologies and prior studies that informed the architecture and validation of the ISMS-CR system.

\bibliographystyle{IEEEtran}

\end{document}